\documentclass{ws-procs975x65}
\usepackage{graphicx}
\usepackage{amsmath}
\usepackage{array}

\begin{document}

\title{On the dynamical instability of self-gravitating systems}
\author{Giuseppe ALBERTI$^*$}
\address{Living Systems Research,\\
Roseggerstra\ss e 27/2, A-9020 Klagenfurt am W\"{o}rthersee, Austria\\
$^*$E-mail: giuseppe.alberti@ilsr.at}

\begin{abstract}
We study the dynamical stability of self-gravitating systems in presence of anisotropy. In particular, we introduce a stability criterion, in terms of the adiabatic local index, that generalizes the stability condition $<\gamma> \geq 4/3$ of the isotropic regime. Also, we discuss some applications of the criterion.
\end{abstract}

\keywords{Anisotropy; Dynamical Instability; Adiabatic Local Index; Gravity.}

\bodymatter

\section{Introduction} \label{sec:1}

\noindent The object of this work consists in the study of the dynamical stability in self-gravitating systems, through the deduction of a stability criterion. In particular, our aim is to obtain a criterion that extends the validity of the stability condition $<\gamma>\geq 4/3$ to the anisotropic systems.

\noindent In Sec. \ref{sec:2} we briefly deduce the stability criterion in Newtonian gravity whereas, in Sec. \ref{sec:3}, we discuss some applications of the criterion in order to quantitatively check how much the presence of the anisotropy can affect the onset of the instability. In Sec. \ref{sec:4} we draw some conclusions.

\section{Stability criterion in Newtonian gravity} \label{sec:2}

\noindent In Newtonian gravity the pulsation equation writes (see Ref.\cite{am_2017, am_2019})

\begin{equation} \label{eq:oscillazioni_newton}
\frac{\partial^2 \xi}{\partial t^2} + \frac{4\xi P'_r}{r\rho} - \frac{1}{\rho} \Bigl[\frac{\gamma_r P_r}{r^2} (r^2 \xi)'\Bigr]' + \frac{6(P_r - P_t) \xi}{r^2\rho} + \frac{2(P_r - P_t) \xi'}{r\rho} + \frac{2(\gamma_t P_t - \gamma_r P_r) (r^2 \xi)'}{r^3\rho} = 0
\end{equation}

\noindent where $P_r$ and $P_t$ are the radial and tangential components of the pressure tensor, respectively, $\rho$ the density, \emph{r} the radial coordinate, $\xi$ the Lagrangian displacement. The adiabatic local indexes $\gamma_r$ and $\gamma_t$, along the radial and the trasverse components, are given by

\begin{equation} \label{eq:gamma}
\gamma_r = \frac{\rho}{P_r} \Big(\frac{\partial P_r}{\partial\rho}\Big)_S \qquad {\rm and} \qquad \gamma_t = \frac{\rho}{P_t} \Big(\frac{\partial P_t}{\partial\rho}\Big)_S \,,
\end{equation}

\noindent where the subscript \emph{S} indicates that the derivatives are performed by keeping the entropy constant. Developing the calculations \cite{am_2019}, we easily get the stability criterion

\begin{equation} \label{eq:criterio_newton}
\mathcal{Q} = \frac{\int_{\,0}^{\,R}(\gamma_r P_r + 2\gamma_t P_t\,)r^2 dr}{\int_{\,0}^{\,R}(P_r + 2P_t)\,r^2 dr}\geq \frac{4}{3} \,.
\end{equation}

\noindent In the isotropic limit the foregoing expression reduces to $\mathcal{Q} = <\gamma> \geq 4/3$.

\begin{figure} \label{fig:politropiche}
\begin{center}
\includegraphics[scale=0.31]{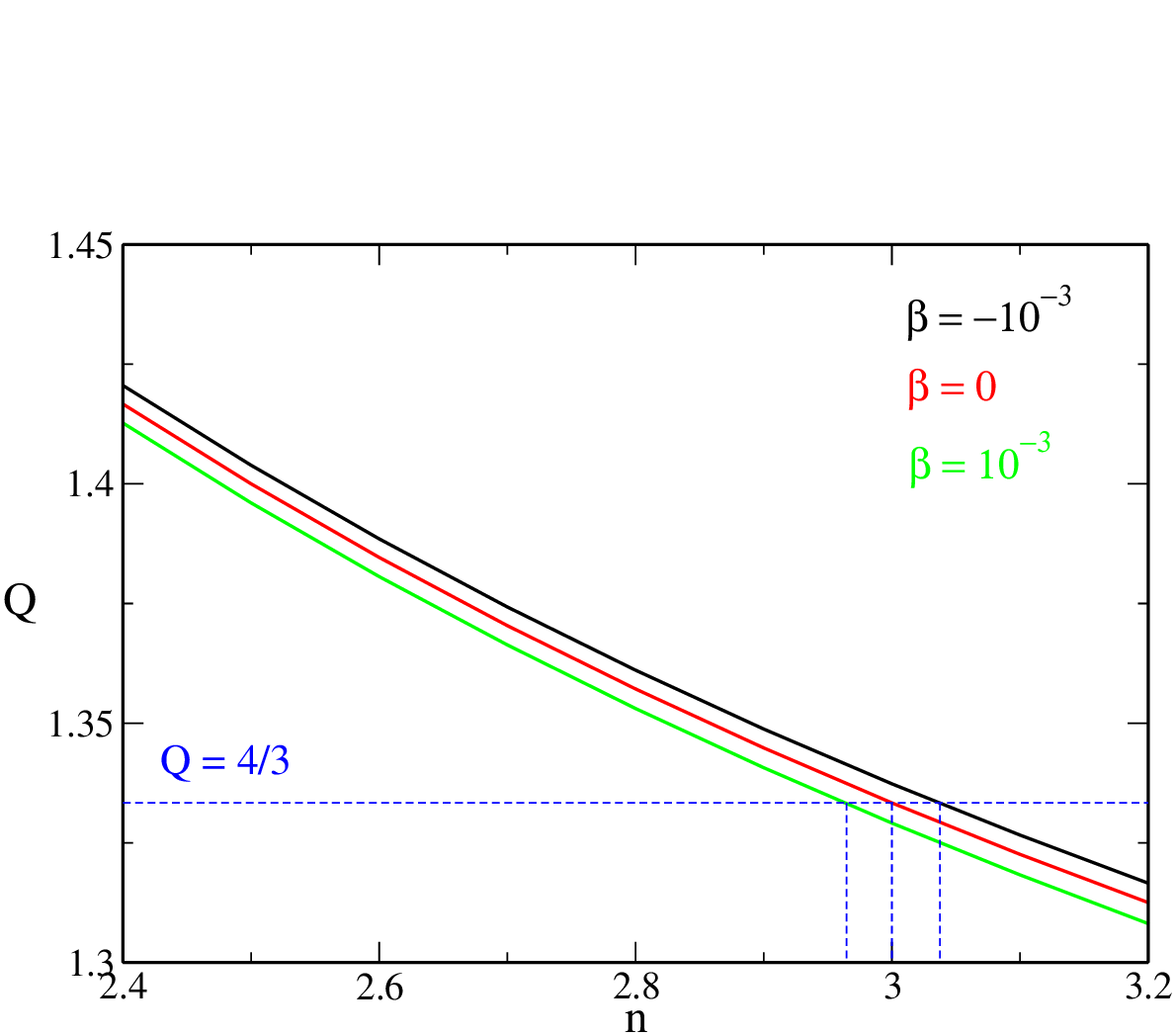}
\includegraphics[scale=0.31]{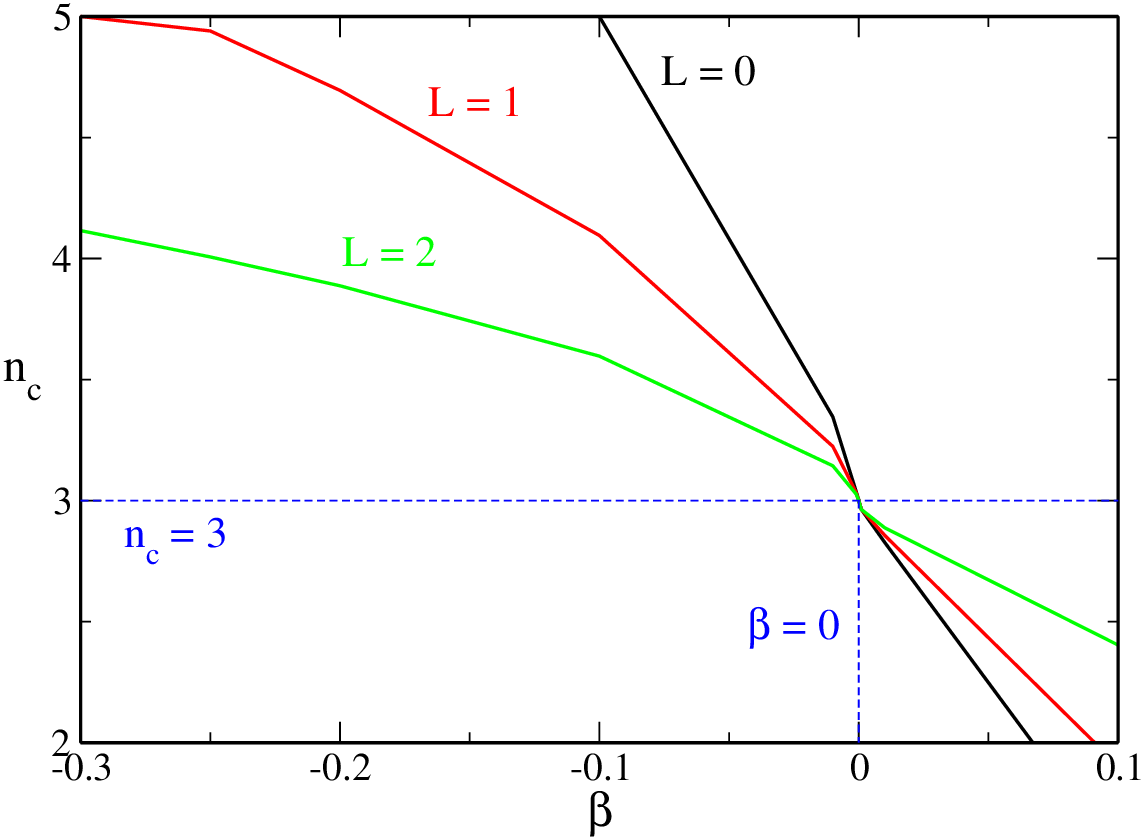}
\caption{\textbf{Left panel}: representation of the function $\mathcal{Q} = \mathcal{Q}(n)$ [see Eq.\eqref{eq:criterio_newton}], for $N = 2$, $L = 0$, $\beta = -10^{-3}$ (black line), 0 (red line) and $10^{-3}$ (green line). The plot shows that an unstable configuration in the isotropic limit can become stable if $\beta < 0$. Vice versa, when $\beta > 0$, a stable configuration in the isotropic limit can become unstable. \textbf{Right panel}: Critical value of the polytropic exponent $n_c$ as a function of the anisotropy parameter $\beta$, for $L = 0$, 1 and 2.}
\end{center}
\end{figure}

\section{Applications} \label{sec:3}
\subsection{Polytropes} \label{subsec:3a}

\noindent Let us apply Eq.\eqref{eq:criterio_newton} to the study of the stability of the polytropic models advanced by Herrera \& Barreto \cite{hb_2013}. Leaving out the details of the calculations\cite{am_2019}, the anisotropic Lane-Emden equation writes

\begin{equation} \label{eq:le_anisotropo}
\frac{1}{\xi^2} \frac{d}{d\xi} \Big(\xi^2 \frac{d\theta}{d\xi}\Big) = -\theta^n \mathcal{F}(\beta) \,,
\end{equation}

\noindent where \emph{n} is the polytropic exponent, $\beta$ the (dimensionless) anisotropy parameter and $\mathcal{F}(\beta)$\footnote{To keep the decreasing behavior of $\theta$, we require that $\mathcal{F}(\beta) > 0$. This condition allows an upper limit for the anisotropy parameter $\beta$ (see Ref.\cite{am_2019}).} is given by

\begin{equation} \label{eq:le_Fbeta}
\mathcal{F}(\beta) = 1 - \frac{\beta}{\xi^2 \theta^n} \frac{d}{d\xi} \Big[f(\xi) \xi^{N+1}\Big] \,.
\end{equation}

\noindent In the foregoing expression, the exponent \emph{N} and the function \emph{f} depend on the model. In the following, we consider $f(\xi) = (1 + \xi)^L$ (with $L \in \mathbb{R}$). The isotropic Lane-Emden equation is recovered for $\beta = 0$ (implying $\mathcal{F}=1$).

\noindent In Fig. 1 (left panel) we have represented the $\mathcal{Q}$-factor [see Eq.\eqref{eq:criterio_newton}] as a function of the polytropic exponent \emph{n}. The diagram shows dissimilar behaviors according to the sign of the anisotropy parameter.

\noindent If $\beta < 0$, which corresponds to the radial anisotropy (i.e. $P_t < P_r$, see Ref.\cite{am_2019}), we observe a tendency towards the stability because the critical value $n_c$ of the polytropic exponent for the onset of the instability is larger than $n = 3$.\footnote{$n = 3$ corresponds to the critical value in the isotropic case.} On the other hand, for $\beta > 0$, which corresponds to the tangential anisotropy (i.e. $P_t > P_r$), we observe that the rising of the dynamical instability is favored (from the plot, we note that $n_c < 3$).

\noindent In the right panel of Fig. 1 we have represented $n_c$ as a function of $\beta$, for three values of the exponent \emph{L}. The plot shows that $n_c > 3$ for $\beta < 0$, in confirmation of the fact that the (presence of) radial anisotropy leads the system to the stability. If $\beta > 0$, by contrast, we find $n_c < 3$, in confirmation of the fact that the tangential anisotropy favors the rising of the instability.

\subsection{Anisotropic stars} \label{subsec:3b}

\noindent In this section we consider two models, advanced by Dev \& Gleiser\cite{dg_2003}, conceived as deviations from the homogeneous model (i.e. $\rho = \rho_0$). In formulae

\begin{subequations} \label{eq:modello_dg1}
\begin{align}
& P_t = P_r + C \rho r^2 \,, \label{subeq:modello_dg1_1} \\
& P_t = P_r (1 + C \rho r^2) \,. \label{subeq:modello_dg1_2}
\end{align}
\end{subequations}

\noindent In the foregoing expressions, \emph{C} estimates the strength of the anisotropy and can be both positive and negative, \emph{a priori}. Similarly to the case of polytropes, $C > 0$ corresponds to the tangential anisotropy and $C < 0$ to the radial anisotropy. Integrating the equilibrium equations by using Eqs.\eqref{subeq:modello_dg1_1} and \eqref{subeq:modello_dg1_2}, we obtain\cite{dg_2003}

\begin{subequations} \label{eq:modello_dg2}
\begin{align}
& P_r = \rho_0^2 \Big(\frac{2\pi G}{3} - C\Big) (R^2 - r^2) \,, \label{subeq:modello_dg2_1} \\
& P_r = \frac{2\pi G\rho_0}{3C} \Big[1 - e^{C\rho_0(r^2 - R^2)}\Big] \,. \label{subeq:modello_dg2_2}
\end{align}
\end{subequations}

\begin{figure} \label{fig:dg_03}
\begin{center}
\includegraphics[scale=0.39]{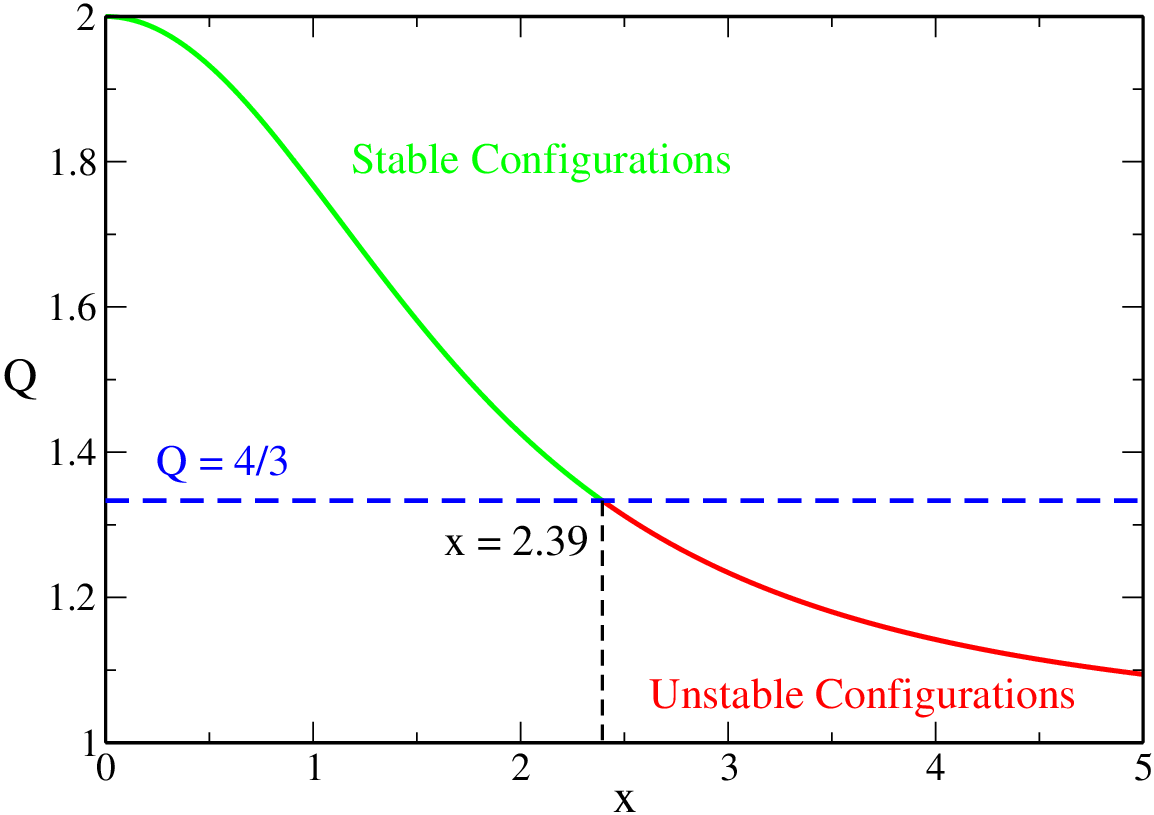}
\end{center}
\caption{Representation of the function $\mathcal{Q} = \mathcal{Q}(x)$, where \emph{x} is defined as $x = R\sqrt{C\rho_0}$. The plot shows the existence of a critical value $x = \bar{x}$, corresponding to the onset of the dynamical instability. Numerically we find $\bar{x} = 2.39$ or, equivalently, $\bar{C} = 5.73/(\rho_0 R^2)$.}
\end{figure}

\noindent In the previous equations, \emph{R} and \emph{G} represent the radius of the star and the gravitational constant, respectively. Concerning the stability, for the ansatz \eqref{subeq:modello_dg1_1}-\eqref{subeq:modello_dg2_1}, Eq.\eqref{eq:criterio_newton} yields $\mathcal{Q} = 2 > 4/3$, showing that the system is dynamically stable. For the ansatz \eqref{subeq:modello_dg1_2}-\eqref{subeq:modello_dg2_2}, conversely, we find a more interesting situation. Eq.\eqref{eq:criterio_newton}, indeed, takes the form

\begin{equation} \label{eq:dg03_criterion_ansatz2}
\mathcal{Q} = 1 + \frac{5}{2x^2} - \frac{15}{4x^4} + \frac{15 D_+(x)}{4x^5} \geq \frac{4}{3} \,,
\end{equation}

\noindent where $x = R\sqrt{C\rho_0}$ and $D_+(x)$ is the Dawson function. In the limit $x \rightarrow 0$, Eq.\eqref{eq:dg03_criterion_ansatz2} reduces to $\mathcal{Q} = 2 > 4/3$ that corresponds to the case previously analyzed. On the other hand, for $x \rightarrow +\infty$, we obtain $\mathcal{Q} = 1 < 4/3$ that corresponds to a loss of stability. Consequently, we expect to find a critical value of \emph{x} allowing the separation between stable and unstable configurations.

\noindent In Fig. 2 we have represented Eq.\eqref{eq:dg03_criterion_ansatz2} and, as we see, the existence of this critical value is confirmed. Numerically, we find that dynamical instability sets in if $x \geq \bar{x} = 2.39$, i.e. if $C \geq \bar{C} = 5.73/(\rho_0 R^2)$. It is interesting to notice that the stability of the star, rather than the anisotropy parameter, depends on the central density.

\subsection{Degenerate fermionic configurations} \label{subsec:3c}

\noindent In this section we focus on the degenerate fermionic configurations, a particular case of a more general study carried out in Ref.\cite{ma_2014}. The equation of state (EOS), in a parametric form, is

\begin{subequations} \label{eq:fermioni_eos}
\begin{align}
& \rho = \frac{4\pi gm^4 \sigma^3}{3h^3} \sqrt{W^3} \Big(1 + \frac{2r^2}{5r_a^2} W\Big) \,, \label{subeq:fermioni_eos_1} \\
& P_r = \frac{4\pi gm^4 \sigma^5}{15h^3} \sqrt{W^5} \Big(1 + \frac{2r^2}{7r_a^2} W\Big) \,, \label{subeq:fermioni_eos_2} \\
& P_t = \frac{4\pi gm^4 \sigma^5}{15h^3} \sqrt{W^5} \Big(1 + \frac{4r^2}{7r_a^2} W\Big) \,. \label{subeq:fermioni_eos_3}
\end{align}
\end{subequations}

\noindent In the foregoing expressions, \emph{W} is the cutoff energy (Fermi energy in this case\cite{ma_2014}), $r_a$ the anisotropy radius, $\sigma$ the velocity dispersion and the other symbols have their usual meaning. The adiabatic local indexes $\gamma_r$ and $\gamma_t$ are given by

\begin{figure} \label{fig:fermioni}
\begin{center}
\includegraphics[scale=0.39]{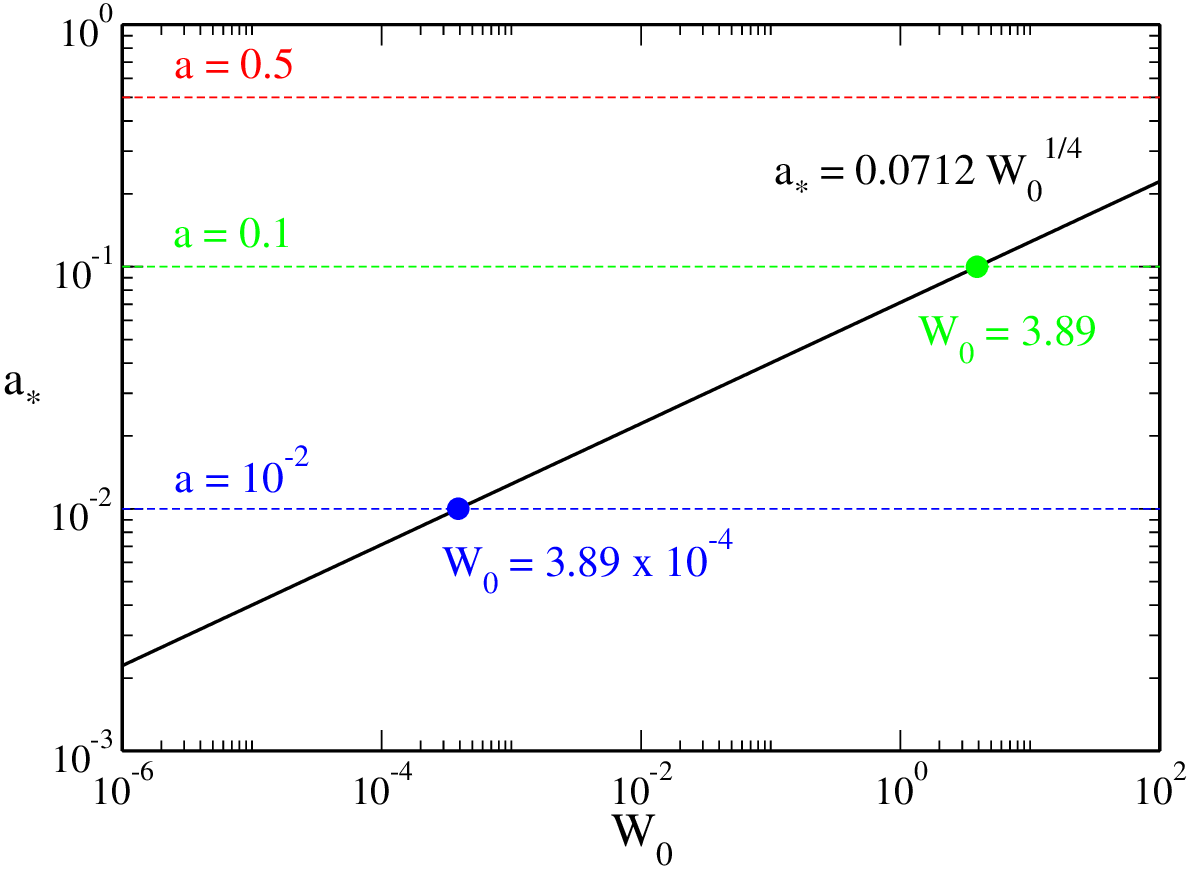}
\end{center}
\caption{Values of $a_*$ as a function of the central density $W_0$ [see Eq.\eqref{subeq:fermioni_eos_1}]. The plot shows that, for a given value of \emph{a} (horizontal lines), there exists a critical value of the central density $W_0^c$ such that dynamical instabilities set in. For $a = 0.5$ (red line) we see that $W_0^c >> 1$ (the system is isotropic and thus stable) whereas, for a $a = 0.1$ (green line), we find $W_0^c = 3.89$ and for $a = 0.01$ (blue line) we have $W_0^c = 3.89 \times 10^{-4}$. This shows that $W_0^c \rightarrow 0$ for $a \rightarrow 0$ (the system is unstable).}
\end{figure}

\begin{subequations} \label{eq:gamma_fermioni}
\begin{align}
& \gamma_r = \frac{5r_a^2 + 2r^2 W}{7r_a^2 + 2r^2 W} \, \frac{8rW^2 + 7W'(5r_a^2 + 2r^2W)}{8rW^2 + 5W'(3r_a^2 + 2r^2W)} \,, \label{subeq:gamma_fermioni_1} \\
& \gamma_t = \frac{5r_a^2 + 2r^2 W}{7r_a^2 + 4r^2 W} \, \frac{16rW^2 + 7W'(5r_a^2 + 4r^2W)}{8rW^2 + 5W'(3r_a^2 + 2r^2W)} \,, \label{subeq:gamma_fermioni_2}
\end{align}
\end{subequations}

\noindent where $W' = dW/dr$. In the isotropic limit $r_a \rightarrow +\infty$ the foregoing expressions yield $\gamma_t = \gamma_r = 5/3$. Consequently, according to Eq.\eqref{eq:criterio_newton}, we get $\mathcal{Q} = 5/3 > 4/3$ confirming that, in the isotropic limit, degenerate fermionic configurations are stable\footnote{We can also show that the EOS can be written as $P_t = P_r = P = K \rho^{5/3}$. This is the EOS of a polytrope of index $n = 3/2$ that, as it is known, represents a stable configuration (being $n = 3/2 < 3$).}. Things change in the fully anisotropic regime $r_a \rightarrow 0$: we have indeed

\begin{equation} \label{eq:gamma_fermioni_anisotropo}
\gamma_t = \gamma_r = \frac{4 + 7y}{4 + 5y} \,, \quad \mbox{ where } \qquad y = \frac{d\log(W)}{d\log(r)} \,.
\end{equation}

\noindent As the reader can see, if $-4/5 < y < -4/7$, $\gamma_r$ and $\gamma_t$ are negative. In these conditions, the speed of the two sound waves along the radial and the tangential axis, given by

\begin{equation} \label{eq:fermioni_suono}
c_{sr} = \sqrt{\frac{\gamma_r P_r}{\rho}} \,, \quad c_{st} = \sqrt{\frac{\gamma_t P_t}{\rho}} \,.
\end{equation}

\noindent become a complex number: the configuration, therefore, is unstable. This feature is a consequence of the trend of density and pressure profiles, which have become ``hollow"\cite{ma_2014}.

\noindent Hollow configurations are characterized by the presence of a maximum achieved far from the center. Consequently, the profile is monotonic increasing until the maximum and monotonic decreasing from the maximum until the boundary.

\noindent A profile is hollow (and thus unstable) if, for a given value of $W_0$, the value of the anisotropy radius $r_a$ is below a critical one. To get this critical value, we compute $d^2\rho/dr^2 = 0$ at the center and solve for \emph{a}. Defining the anisotropy parameter as $a = r_a/r_0$ ($r_0$ is a scaling length\cite{ma_2014}) and indicating by $a_*$ the critical value, we have

\begin{equation} \label{eq:fermioni_astar}
a_* = \frac{r_a}{r_0}\Big|_{critical} = \frac{\sqrt[4]{W_0}}{2\pi \sqrt{5}} \,.
\end{equation}

\noindent In Fig. 3 we have represented the function $a_* = a_* (W_0)$. According to the plot, degenerate fermionic configurations are stable if, for a given value of $W_0$, $a > a_*$ and unstable if $a \leq a_*$.

\section{Concluding remarks} \label{sec:4}

\noindent In this work we have studied the dynamical stability of anisotropic self-gravitating systems. The analysis carried out has shown that, according to the type of anisotropy, the onset of the instability is modified.

\noindent In prevalence of radial anisotropy ($P_r > P_t$), we have observed that the systems have the tendency to evolve towards stable configurations. Unstable configurations in the isotropic limit ($P_r = P_t$) can become stable. In prevalence of tangential anisotropy ($P_r < P_t$), by contrast, we have observed that the rising of the instability is favored. Stable configurations in the isotropic limit can become unstable.

\noindent In Ref.\cite{am_2019} we have studied the stability of other systems, such as Elliptical Galaxies.

\noindent We think that the stability criterion introduced and applied in this work represents a powerful tool for the study of the dynamical stability of a large class of astrophysical systems. The extension of the criterion to General Relativity and other applications will be adressed to forthcoming publications.

\end{document}